\documentstyle[]{article}
\title{On the Significance of the Fifth Coordinate in Wesson's Version of Kaluza-Klein Theory}
\author{N. Redington\\Building 14S-100\\Massachusetts Institute of Technology\\Cambridge, MA 02139, USA}
\date{}
\begin {document}
\maketitle
\begin{center}
\begin{it}
ABSTRACT:
\end{it} \\

\begin{it}
It is argued that the fifth coordinate should correspond to an intensive 
parameter rather than to rest-mass as originally proposed by Wesson.
\end{it}
\end{center}

Traditionally, the fifth coordinate in Kaluza-Klein theories has been 
considered unobservable. This ``cylindricity condition'' has been justified 
on a number of grounds, especially on that of dimensional compactification. 
In the last decade, however, Wesson and his colleagues $^{1-3}$ have argued 
that five-dimensional general relativity without matter and without the 
cylindricity condition is equivalent to normal four-dimensional relativity if 
all the explicitly $x^{5}$ dependent terms are equated to the four-dimensional
stress tensor. Although this theory does not require any specific 
interpretation of $x^{5}$, Wesson {\it et al.} tentatively identify the fifth 
dimension as rest mass.

This suggestion, which was also made independently by Neacsu$^{4}$ in 1981, 
seems to me to face an insuperable difficulty. Mass is additive almost by 
definition. In Newtonian mechanics, two separated objects may be regarded as 
a single entity at some average location, with a mass equal to the sum of the 
constituent masses. This property is retained, at least in the classical 
limit, both in relativity and in quantum mechanics. 

By contrast, space and time coordinates are not additive for separated events:
a mountain range cannot be replaced by a single peak as high as a hundred 
mountains, nor are two events in the Middle Ages conflatable into one event 
now! Any physical quantity which is to be identified with a coordinate must 
not, in the classical limit, behave like an extensive variable. This would 
seem  to eliminate both the rest-mass interpretation and the proposal of 
Ruder$^{5, 6}$ that $x^{5}$ in the compactified Kaluza theory be identified 
with action.

If $x^{5}$ is not an extensive variable, it is natural to ask whether any 
intensive variable suggests itself as a candidate. One fairly obvious 
possibility is mass {\it density}, which would make mass into a kind of 
4-volume and therefore additive. In such an interpretation, point masses of 
density $\rho = m \delta (x^{i} - x^{i}_0)$ would appear rather like long 
tears in the fabric of space time. 

Such an identification makes the relationship between the time-time component 
of the stress tensor and the fifth dimension slightly more tractable than in 
the rest-mass interpretation. For a pure fluid, the trace of the 
five-dimensional Ricci tensor in Wesson's theory is$^{7}$:
 $$ (8 \pi G / c^{4})(\rho c^{2} - 3 P) = (1 / 4 g_{55})(\partial_{5}g^
{\alpha \beta} \partial_{5}g_{\alpha \beta} + (g^{\alpha \beta} 
\partial_{5}g_{\alpha \beta})^{2})    $$ 

In the weak field limit at low pressure, this reduces to 
$$ \eta^{\alpha \beta} h_{\alpha \beta},_{55} = (32 \pi G / c^{2}) \rho   $$ 

So if $x^{5}$ is proportional to $\rho$, 
$$h_{\alpha}^{\alpha} \sim (16 \pi G / 3c^{2}) \rho^{3} + b(x^{\mu}) 
\rho + a(x^{\mu})$$ 

Of course, there is no reason to prefer density over any other intensive 
variable. Temperature, for instance, would work just as well; such an 
interpretation might provide hope for a geometrization of thermodynamics. 
Many of these possibilities are moreover experimentally testable; unlike 
rest-mass, both density and temperature can be varied in the laboratory. It 
is interesting that a number of unusual and unexplained phenomena, such as 
sonoluminescence$^{8}$, involve sudden large changes in temperature and/or 
density.   

Finally, it is perhaps worth recalling that, in the original Kaluza theory, 
electric charge is the fifth component of momentum.  The coordinate $x^{5}$ 
should then in that theory be identified with the quantity conjugate to charge
 in the uncertainty relations. That quantity is magnetic flux.$^{9}$

\begin{center}
{\bf BIBLIOGRAPHY}
\begin{enumerate}
 
\item P. S. Wesson 1990 {\it Gen. Rel. \& Grav.} 22(707)

\item P. S. Wesson \& J. Ponce de Leon 1992 {\it J. Math. Phys.} 33(3883)

\item P. S. Wesson {\it et al.} 1996 {\it Int. J. Mod. Phys.} A11(3247)

\item M. A. Neacsu 1981 M.S. Thesis, Texas Tech University.

\item Yu. B. Rumer 1949 {\it Zh. Eksp. Teor. Fiz.} 19(86 and 207)

\item Yu. B. Rumer 1959 {\it Zh. Eksp. Teor. Fiz.} 36(1894) [{\it Sov. Phys. JETP} 36(1348)]

\item J. M. Overduin \& P. S. Wesson 1997 {\it Phys. Rep.,} (forthcoming); http://astro.uwaterloo.ca/$\sim$ wesson/\* PUB

\item L. A. Crum and R. A. Roy 1994 {\it Science} 266(233)

\item A. Widom {\it et al.} 1985 {\it Phys. Rev.} B 31(6588)

\end{enumerate}
\end{center} 
 
\end{document}